\documentclass{pasj02}
\usepackage{amsmath}
\usepackage[switch,mathlines]{}
\Received{$\langle$reception date$\rangle$}
\Accepted{$\langle$acception date$\rangle$}
\Published{$\langle$publication date$\rangle$}

\begin{document}

\title{Near-Infrared Photometry of SN 2023bee and the Standardization of Type Ia Supernovae}
\author{Masayuki Yamanaka$^{1}$ \& Takahiro Nagayama$^{2}$}

\altaffiltext{1}{Amanogawa Galaxy Astronomy Research Center (AGARC), Graduate School of Science and Engineering, Kagoshima University, 1-21-35 Korimoto, Kagoshima, Kagoshima 890-0065, Japan}
\email{yamanaka@sci.kagoshima-u.ac.jp}

\altaffiltext{2}{Graduate School of Science and Engineering, Kagoshima University, 1-21-35 Korimoto, Kagoshima, Kagoshima 890-0065, Japan}


\maketitle

\begin{abstract}
We report near-infrared (NIR) photometric observations of the slow-declining Type Ia supernova (SN Ia) 2023bee, spanning the phase from $t=-12$ to $t=93$ d. 
At the first maximum, the absolute magnitudes reached $-18.7$, $-18.4$, and $-18.6$ mag in the $J$, $H$, and $K_{\rm s}$ bands, respectively. 
SN 2023bee is brighter than the mean of normal SNe Ia.
The $J$ and $H$-band light curves are highly consistent with those of the shallow-silicon SN 2012fr.
A statistical analysis was performed using data for SN 2023bee and samples collected from the literature.
 We found that the first maximum absolute magnitudes are correlated with those of the secondary maximum in the $H$ 
($r = 0.8657$; $p = 0.0003$) and $K_{s}$ ($r = 0.8996$; $p = 0.0004$) bands. 
 We found a strong correlation in the $J$ band between the first maximum absolute magnitude and the time interval between the first and secondary maxima ($r = -0.854$; $p = 0.0004$). The corresponding correlation in the $H$ band was weaker ($r=-0.725$; $p = 0.0116$). This implies that these time differences have the potential to serve as a distance indicator for host galaxies. 
\end{abstract}



\section{Introduction}
Type Ia supernovae (SNe Ia) are a powerful tool for constraining cosmological parameters \citep{Riess1998,Perlmutter1999}, owing to the strong correlation between their absolute magnitude and the decline rate of their light curve \citep{Phillips1993,Phillips1999}. 
Although the width-luminosity relation has enabled SNe Ia to be used as powerful distance indicators for three decades, fundamental questions regarding their explosion mechanisms \citep{Hoyle1960,Nomoto1984,Khokhlov1991,Kasen2009}
and progenitor models \citep{Tutukov1979,Nomoto1982,Iben1984}
remain unresolved.


\begin{figure*}
\includegraphics[width=1.0\textwidth]{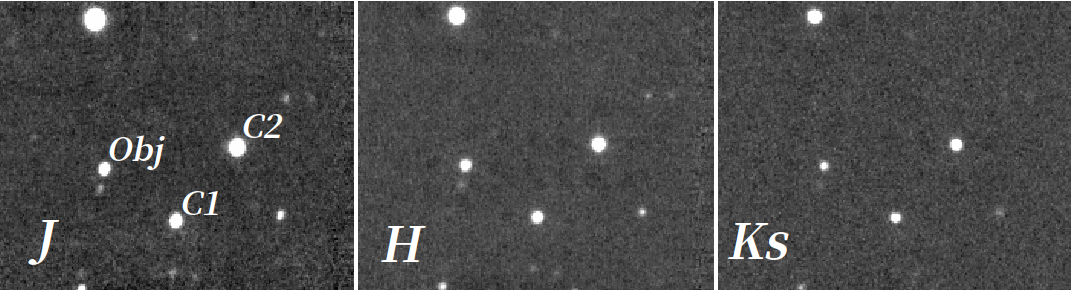}
\caption{Field images of SN 2023bee in the $J$, $H$, and $K_{\rm s}$ bands obtained on 2023 March 28 with the kSIRIUS instrument at the Iriki Observatory 1.0-m telescope. 
Numbers indicate the reference stars used for photometric calibration.
{Alt text: Three monochrome images.}}, 
\end{figure*}

Luminous SNe Ia generally exhibit slow-declining light curves \citep{Phillips1992}. Their spectra typically show relatively high-excitation Fe and shallow silicon absorption features, resulting in their classification as SN 1991T/SN 1999aa-like objects \citep{Filippenko1992b,Branch1993,Garavini2004}.
These objects have a small depth ratio between the two Si~{\sc ii} absorption lines, and this ratio is inversely proportional to the luminosity \citep{Nugent1995,Hatano2000,Hachinger2006}. Moreover, based on Si~{\sc ii} line velocity measurements, this subclass corresponds to SNe Ia with a low velocity gradient \citep{Benetti2005}.
 
Recently, the near-infrared (NIR) light curve has emerged as a powerful distance indicator, primarily because the NIR absolute magnitudes show less dispersion among SNe Ia compared to their optical counterparts \citep{Meikle2000,Krisciunas2004,Wood-Vasey2007,Phillips2012,Dhawan2015,Peterson2023,Do2026}. Consequently, the sophistication of template NIR light curves for use as a luminosity indicator and their application to constraining the cosmological constant have been investigated \citep{Barone-Nugent2012,Jones2022,Mandel2022,Dhawan2023}.
 
NIR light curves of normal and luminous SNe Ia exhibit a pronounced secondary maximum \citep{Elias1981,Elias1985,Wells1994,Sadakane1996,Krisciunas2001,Krisciunas2003}. The luminosity of a SN Ia is thought to be powered by the radioactive decay of the synthesized $^{56}$Ni mass \citep{Arnett1982}. SNe with a brighter maximum exhibit a more luminous and later secondary maximum \citep{Elias-Rosa2008,Dhawan2015}. The observational properties of the secondary peak have been pointed out to be a probe of the radioactive $^{56}$Ni and/or iron-group element masses, the ejecta structures, and the explosion mechanisms \citep{Kasen2006,Jack2015}. Therefore, investigation of the properties of the double-peaked NIR light curve is desirable to discuss the explosion model independent of the optical.
 
 NIR observations serve as a potential probe for illustrating the presence of the circumstellar material around the SN Ia progenitor system \citep{Maeda2015}. In certain peculiar SNe Ia, strong NIR emission originates from the circumstellar material \citep{Yamanaka2016,Nagao2017}. To confirm the NIR excess, more template light curves are necessary. Samples of the long-duration NIR light curves are still limited. Recently, photometric data extended to the extremely late epoch were obtained using the {\it Hubble Space Telescope}, and the light curves definitively unveiled a conspicuous plateau \citep{Graur2020,Deckers2023}. This provided circumstantial evidence of the "infrared catastrophe" scenario \citep{Axelrod1980}, but an exception was found \citep{Graur2023}.
 
 SN 2023bee was discovered at $17.3$ mag on Feb 1 in the nearby galaxy NGC 2708 by the Distance Less Than 40 Mpc Survey (DLT40; \cite{Andrews2023}), and identified as an early-phase SN I \citep{QZhai2023} or a SN Ia \citep{Hosseinzadeh2023b} by the spectroscopic observations. Analysis of the extremely early-phase ultra-violet and optical photometric data unveiled a significant excess potentially originating from shock cooling of a progenitor or circumstellar medium (CSM), and constraints on the explosion environments and models were discussed \citep{Hosseinzadeh2023,QWang2024}. This SN was categorized as shallow silicon based on its spectra \citep{Hosseinzadeh2023,QWang2024}.
 
 This paper focuses on NIR photometric data obtained in our framework and is presented as follows: $\S 2$ describes the observations and data reduction; $\S 3$ presents the light curve and color evolution; the NIR light curve properties are discussed based on a statistical analysis in $\S 4$; and $\S 5$ presents our conclusion.

\section{Observations and Reduction}
 We carried out NIR imaging observations of SN 2023bee in the $J$, $H$, and $K_{\rm s}$ bands on $23$ nights from Feb 4.6 to May 21.5 using the NIR simultaneous three-band camera, kSIRIUS \citep{Nagayama2024}, attached to the 1-m telescope at the Iriki Observatory.

Data reduction was performed according to the standard procedures for NIR imaging data. A template sky image was created by combining the dithered images. This template was then subtracted from each of the individual images to remove the sky background. Finally, all the sky-subtracted images obtained on each night were stacked by taking their mean. Point-spread-function (PSF) model-fitting photometry was then performed using the IRAF/DAOPHOT package \citep{Stetson1987} on the object and the reference stars shown in Figure~1.
Photometric calibration was conducted using these references, based on the Two Micron All Sky Survey catalog (2MASS; \cite{Skrutskie2006}). The obtained $JHKs$-band magnitudes are listed in Table 1. The observed phases of the light curves were corrected for cosmological time dilation using the redshift ($z=0.0067$; \cite{Pisano2002}) of the heliocentric velocity of the host galaxy NGC 2708. The $JHKs$-band light curves are presented in Figure~2. The $B$-band maximum date, MJD 59992.0 \citep{Hosseinzadeh2023,QWang2024}, was adopted as $t=0$ throughout this paper.

\begin{table*}
 \caption{Log of the photometry of SN 2023bee}
    \centering
    \begin{tabular}{llllllll}
    \hline
MJD & Phase & $J$ & err($J$) & $H$ & err($H$) & $K_{\rm s}$ & err($K_{\rm s}$) \\
(day) &     & (mag) &  & (mag) & & (mag) &  \\
    \hline
59979.6 & -12.32 & 15.336 & 0.032 & 15.384 & 0.059 & 15.394 & 0.077 \\
59989.7 & -2.28 & 13.818 & 0.009 & 14.108 & 0.017 & 13.986 & 0.021 \\
59991.7 & -0.30 & 13.885 & 0.009 & 14.168 & 0.019 & 13.976 & 0.015 \\ 
60000.6 & 8.54 & 14.597 & 0.011 & 14.482 & 0.018 & 14.164 & 0.035 \\
60001.6 & 9.54 & 14.740 & 0.016 & 14.467 & 0.025 & 14.338 & 0.028 \\ 
60002.6 & 10.53 & 14.940 & 0.010 & 14.508 & 0.025 & 14.211 & 0.020 \\ 
60008.6 & 16.49 & 15.733 & 0.037 & 14.583 & 0.018 & 14.363 & 0.028 \\ 
60009.6 & 17.48 & 15.745 & 0.013 & 14.509 & 0.020 & 14.365 & 0.021 \\ 
60010.6 & 18.48 & 15.765 & 0.016 & 14.485 & 0.023 & 14.250 & 0.016 \\ 
60016.6 & 24.44 & 15.605 & 0.011 & 14.350 & 0.020 & 14.224 & 0.024 \\
60018.5 & 26.32 & 15.522 & 0.012 & 14.266 & 0.017 & 14.201 & 0.030 \\ 
60022.6 & 30.40 & 15.384 & 0.022 & 14.209 & 0.045 & 14.204 & 0.047 \\ 
60031.5 & 39.24 & 15.204 & 0.013 & 14.468 & 0.015 & 14.701 & 0.027 \\ 
60036.5 & 44.20 & 15.595 & 0.019 & 14.795 & 0.017 & 14.916 & 0.030 \\ 
60042.5 & 50.16 & 16.065 & 0.017 & 15.157 & 0.019 & 15.250 & 0.026 \\ 
60043.5 & 51.16 & 16.156 & 0.019 & 15.123 & 0.023 & 15.300 & 0.024 \\ 
60044.6 & 52.25 & 16.186 & 0.037 & 14.983 & 0.049 & 15.314 & 0.073 \\ 
60051.5 & 59.10 & 16.743 & 0.030 & 15.464 & 0.028 & 15.743 & 0.054 \\ 
60055.6 & 63.18 & 17.071 & 0.044 & 15.673 & 0.043 & 16.055 & 0.121 \\ 
60060.5 & 68.04 & 17.190 & 0.035 & 15.819 & 0.032 & 15.778 & 0.056 \\ 
60065.5 & 73.01 & 17.426 & 0.076 & $\cdot$$\cdot$$\cdot$ & $\cdot$$\cdot$$\cdot$ & $\cdot$$\cdot$$\cdot$  & $\cdot$$\cdot$$\cdot$  \\ 
60073.5 & 80.96 & 17.836 & 0.086 & 16.471 & 0.078 & 16.177 & 0.074 \\ 
60085.5 & 92.88 & 18.383 & 0.117 & 16.845 & 0.089 & 16.519 & 0.156 \\
 \hline
     \end{tabular}
\end{table*}

\begin{figure}
\includegraphics[width=0.45\textwidth]{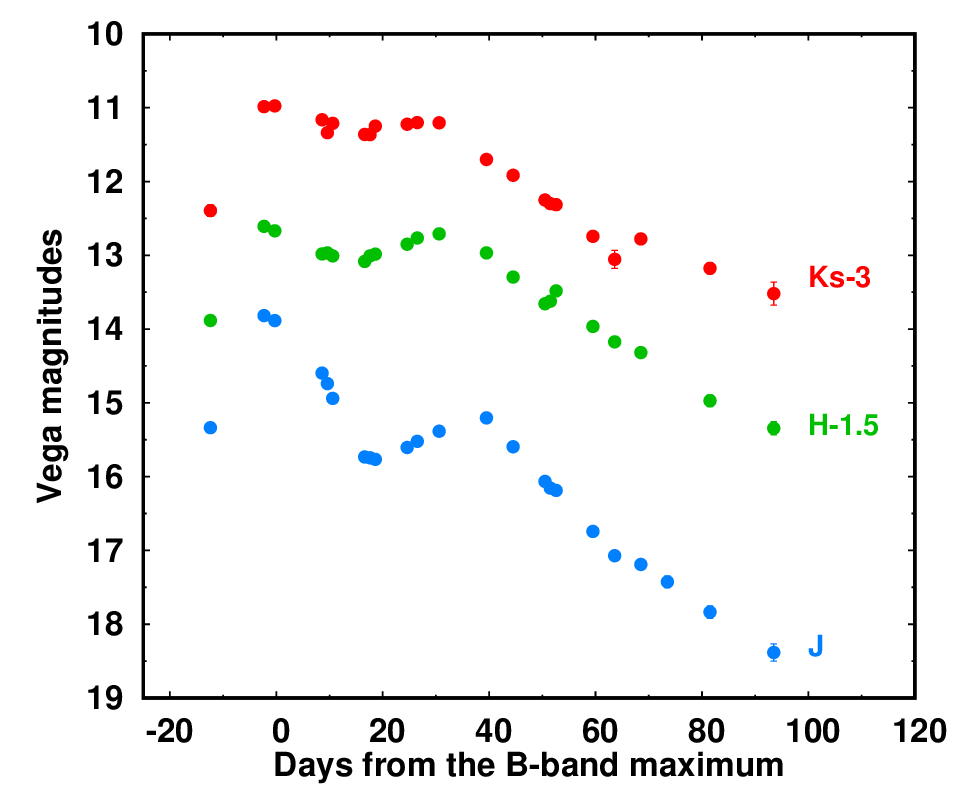}
\caption{The near-infrared (NIR) light curves of SN 2023bee in the $J$, $H$, and $K_{\rm s}$ bands. The data cover the phase until 100 days after the $B$-band maximum date ($t_{\text{max}, B}$). The magnitudes are vertically offset for clarity to distinguish the photometric points in each band. {Alt text: A scatter plot.}}
\label{fig:lightcurve_2023bee}
\end{figure}

\section{Results}
\subsection{Light curves and color evolution}
 The first and secondary maxima were determined by fitting the light curves with polynomial functions (see \S 4 for the detailed description.). 
The first maximum magnitudes in $J$, $H$, and $K_{\rm s}$-bands were 
13.8, 14.1, and 13.9 mag at $t=-1.7$, $-2.4$, and $0.5$ d, 
respectively. The secondary maximum magnitudes were 
15.2, 14.2, and 14.2 mag at $t=36.7$, $30.4$, and $28.6$ d. 
The secondary maximum is observed later in bluer bands, a trend also seen in SNe Ia, while the secondary maximum 
magnitude of the $J$-band was $\sim1.4$ mag fainter 
than the first one. The first maximum magnitudes
were almost similar to the secondary ones in $H$ and $K_{\rm s}$ bands. 
After the secondary maximum, 
the light curves showed a linear decline in the $H$ bands,
while the decline rates in the $J$ and $K_{\rm s}$ bands decreased after $t = 60$ d compared with those at earlier phases.
 
 We obtained the absolute magnitudes of SN 2023bee.
The Galactic extinction was corrected using the extinction values from \citep{Schlafly2011}. The host galactic 
extinction was negligible \citep{Hosseinzadeh2023,QWang2024}.
The distance modulus was adopted as $\mu=32.5$ mag \citep{QWang2024}.
The absolute magnitudes of the first maximum were 
$M_J\sim-18.7$, $M_H\sim-18.4$, and $M_{Ks}\sim-18.6$ mag, and the 
secondary maximum are $M_J\sim-17.3$, $M_H\sim-18.3$, and $M_{Ks}\sim-18.3$ mag. 

\begin{figure*}
\includegraphics[width=1.0\textwidth]{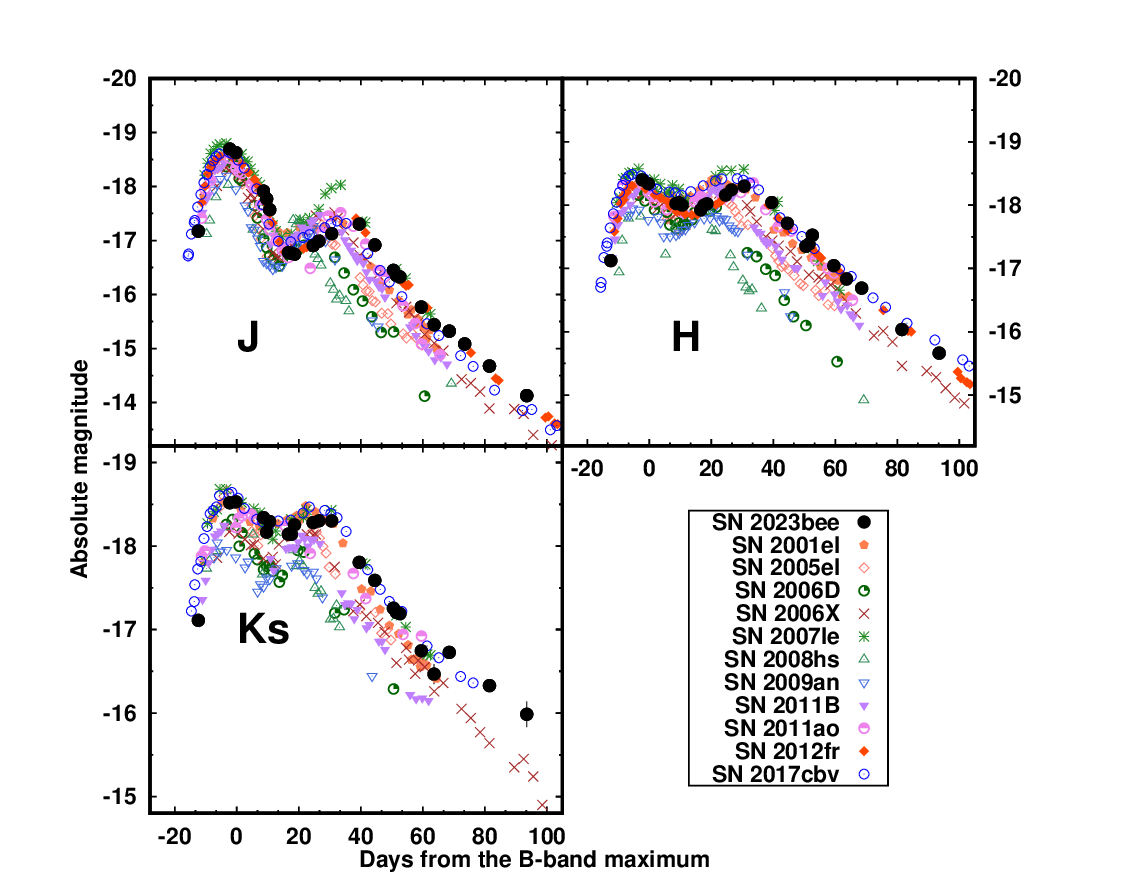}
\caption{The absolute magnitude light curves of SN 2023bee in the $J$, $H$, and $K_{\rm s}$ bands, shown in comparison with a sample of other SNe Ia. The comparison SNe data were adopted from \citet{Krisciunas2003,Friedman2015,Contreras2018,Wee2018}. Corrections for extinction (host and Galactic) and distance moduli were applied using the values published in the corresponding references. The phase is given relative to the $B$-band maximum epoch ($t=0$). {Alt text: A three-panel figure arranged in two columns.}}
\label{fig:abs_lc_comparison}
\label{fig:abs_lc_comparison_short}
\end{figure*}



The $J$, $H$, and $K_{s}$-band absolute magnitude light curves are shown in Figure~3, together with those of similar SNe Ia.
The comparison sample includes SNe 2001el \citep{Krisciunas2003}, 2012fr \citep{Contreras2018}, and 2017cbv \citep{Wee2018}.
Additional data for SNe 2005el, 2006X, 2007le, 2008hs, 2011B, and 2011an were taken from the large samples obtained by \citet{Friedman2015}. 

The criteria we used to select the comparison sample are as follows.
(1) For the comparison of the light curves and color evolution, we selected SNe with observations available in all three bands ($J$, $H$, and $K_s$). This criterion was also applied to our statistical analysis, except for SN 2012fr, as described below. (2) The phase was defined relative to the $B$-band maximum, whose epoch was determined from optical light-curve studies reported in the literature. (3) In each of the $J$, $H$, and $K_s$ bands, the observations covered a phase 
range extending from before $t=-3$ d to beyond $t=+50$ d. (4) In 
each of the $J$, $H$, and $K_s$ bands, the interval between any two consecutive observations was no longer than 5 days.
One important exception is SN 2012fr, which is very similar to SN 2023bee. We included this object in the statistical analyses involving the $J$ and $H$ bands, despite the lack of $K_s$-band data. Accordingly, SN 2012fr was excluded from analyses involving the $K_s$ band.

We also considered samples from \citet{Hamuy2006}, \citet{Johansson2021}, and \citet{Peterson2023}. 
However, these do not satisfy all of our criteria. 
The samples of \citet{Johansson2021} and \citet{Peterson2023} present $YJH$-band data but lack $K_s$-band observations. Although the CSP sample includes $K_s$-band, the pre-maximum data are insufficient for our criteria. 
In addition, the cadence of the available light curves does not satisfy the criteria. Therefore, we did not include these samples in our comparison sample or statistical analysis.

The host-galaxy extinction values and distance moduli for all comparison SNe Ia were adopted from the corresponding references \citep{Krisciunas2003,XWang2008,Simon2009,Scalzo2014,Dhawan2017,Contreras2018,Wee2018,Avelino2019}.
The $B$-band maximum date, adopted as $t=0$, was taken from the above studies.
The first-maximum magnitudes of SN 2023bee are 0.2 mag brighter than the average values for the comparison SNe Ia in the $J$, $H$, and $K_s$ bands, while the secondary-maximum magnitudes are 0.1 mag brighter.


The $J$-band light curves showed similar morphologies around the first maximum among all SNe.
Excluding SNe 2008hs, 2007le, and 2009an, the absolute magnitudes at the secondary maximum were comparable across the remaining sample, with a standard deviation of 0.05 mag.
However, after the secondary maximum, the onset of the final decline showed a large deviation in time, although their subsequent decline rates were similar.

In the $H$ and $K_{\rm s}$ bands, the absolute magnitudes of the first maximum exhibited large diversity.
The comparison sample showed a standard deviation of $\sim0.7$ mag in the $H$-band and $\sim0.8$ mag in the $K_{\rm s}$-band at maximum light.
The decline rates of the light curves after the secondary maximum were similar in the $K_{\rm s}$ band, but this similarity was not observed in the $H$ band.

The overall light curve morphology of SN 2023bee was similar to those of SNe 2012fr and 2017cbv.
Specifically, the $J$ and $H$-band light curves were quite similar to those of SN 2012fr.
After $t=60$ d, the $K_{\rm s}$-band light curve evolved more slowly than in the previous phase, a trend also similar to that seen in SN 2017cbv.
SNe 2012fr and 2017cbv have been classified as shallow-silicon SNe Ia based on the equivalent widths of Si~{\sc ii} \citep{Childress2013,Burns2020}, and their optical light curve evolution is known to be slow \citep{Contreras2018,Wee2018}.
The optical spectrum and light curve of SN 2023bee are also consistent with this subclass designation \citep{Hosseinzadeh2023,QWang2024}.
 
\subsection{Near-infrared color evolution}

Figure~4 shows the $J-H$ and $J-Ks$ color evolution compared with those of other SNe Ia. The comparison SNe are the same as those used in Figure~3. 
The host-galaxy reddening, $E(B-V)$, was adopted from the literature, where it had been estimated by comparing the observed color evolution with template colors, using either the Lira relation \citep{Lira1998} or the color at the $B$-band maximum \citep{Phillips1999}.
The $J-H$ color evolved to reach a red peak at $t=20$ d.
This timing of the red peak was as late as those observed for SNe 2012fr and 2017cbv.
The overall evolution closely followed that of SN 2012fr. Up to $t=30$ d, the color evolution was also comparable to that of SN 2017cbv.
After $t=30$ d, SNe 2023bee and 2012fr evolved toward bluer colors.

In $J-K_{\rm s}$, a similar trend was observed. The color became almost the same as that of SN 2017cbv after $t=60$ d.
The slow evolution is consistent with SNe that exhibit slow-declining light curves. At $t=60$ d, the $J-K_{\rm s}$ color suddenly reddened.
Interestingly, the $J-K_{\rm s}$ color evolution was similar 
to that of SNe 2006X and 2017cbv.

SN 2023bee showed one of the bluest colors among the sample SNe Ia during $t=-10$-$10$ d and $40$-$60$ d.
Generally, the reddening effect on NIR color is not as large as that on optical color.
This fact supports the conclusion that the host galactic extinction is almost negligible, which is consistent with previous works \citep{Hosseinzadeh2023,QWang2024}.

\begin{figure*}
\includegraphics[width=1.0\textwidth]{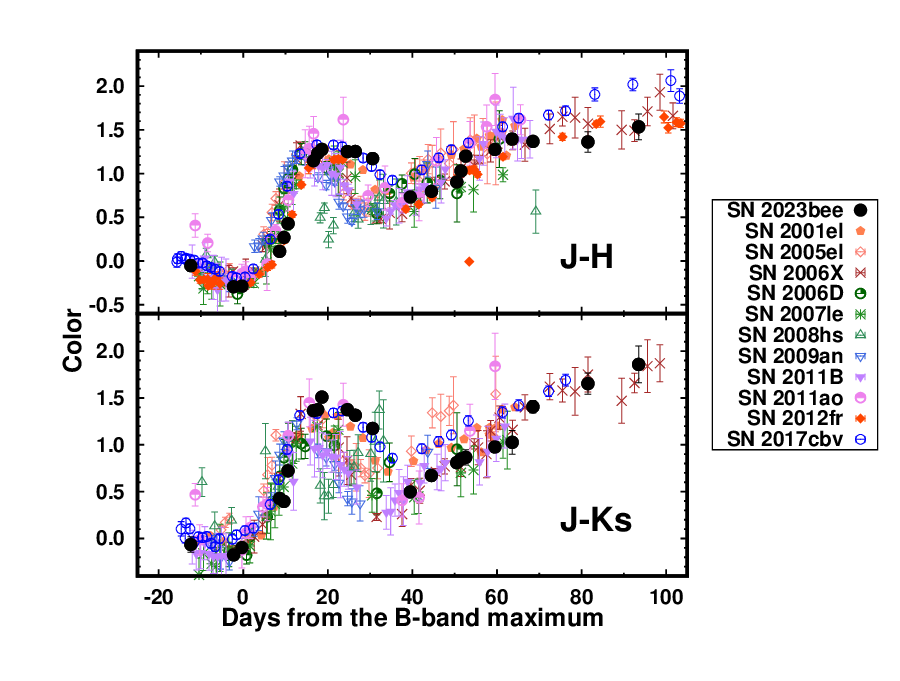}
\caption{The $J-H$ and $J-K_{\rm s}$ color evolution compared with those of a select sample of other SNe Ia. 
The samples used for comparison are represented by symbols with the same shapes and colors as those used in Figure 3.
{Alt text: A two-panel figure arranged in three lines.}}
\end{figure*}

 \begin{table*}
     \centering
\caption{Summary of the absolute magnitudes at the first ($M_{\lambda, 1st}$) and secondary ($M_{\lambda, 2nd}$) maxima for SN 2023bee and the comparison SNe in the $J$, $H$, and $K_{\rm s}$ bands.}
     \begin{tabular}{lllllllllllll}
     \hline
Object & $M_{J,1st}$ & err & $M_{H,1st}$ & err &$M_{Ks,1st}$ & err &$M_{J,2nd}$ & err &$M_{H,2nd}$ & err &$M_{Ks,2nd}$ & err  \\
 & (mag) &  & (mag) &   & (mag) & & (mag) & & (mag) &  & (mag) &  \\

     \hline \hline
23bee & -18.68 & 0.01 & -18.39 & 0.01 & -18.56 & 0.05 & -17.31 & 0.04 & -18.30 & 0.00 & -18.32 & 0.02 \\ 
01el & -18.62 & 0.02 & -18.31 & 0.03 & -18.54 & 0.03 & -17.42 & 0.05 & -18.38 & 0.01 & -18.43 & 0.03 \\ 
05el & -18.45 & 0.02 & -18.35 & 0.06 & -18.62 & 0.07 & -17.33 & 0.04 & -18.19 & 0.03 & -18.26 & 0.02 \\ 
06D & -18.35 & 0.00 & -18.18 & 0.05 & -18.15 & 0.13 & -16.88 & 0.03 & -17.95 & 0.07 & -17.94 & 0.04 \\ 
06X & -18.35 & 0.02 & -18.14 & 0.00 & -18.21 & 0.01 & -17.52 & 0.06 & -18.22 & 0.00 & -18.17 & 0.02 \\ 
07le & -18.83 & 0.02 & -18.55 & 0.01 & -18.68 & 0.03 & -18.05 & 0.01 & -18.58 & 0.01 & -18.46 & 0.04 \\ 
08hs & -18.05 & 0.04 & -17.89 & 0.05 & -18.35 & 0.00 & -17.47 & 0.06 & -17.99 & 0.03 & -17.95 & 0.09 \\ 
09an & -18.18 & 0.03 & -17.92 & 0.02 & -17.98 & 0.03 & -17.14 & 0.02 & -17.82 & 0.00 & -17.79 & 0.00 \\ 
11B & -18.34 & 0.08 & -18.19 & 0.01 & -18.31 & 0.04 & -17.40 & 0.09 & -18.22 & 0.01 & -18.07 & 0.01 \\ 
11ao & -18.48 & 0.01 & -18.34 & 0.03 & -18.36 & 0.01 & -17.53 & 0.12 & -18.25 & 0.01 & $\cdot\cdot\cdot$ & $\cdot\cdot\cdot$ \\ 
12fr & -18.67 & 0.03 & -18.35 & 0.04 & $\cdot\cdot\cdot$ & $\cdot\cdot\cdot$ & -17.45 & 0.05 & -18.18 & 0.04 & $\cdot\cdot\cdot$ &  $\cdot\cdot\cdot$ \\ 
17cbv & -18.66 & 0.01 & -18.48 & 0.04 & -18.63 & 0.01 & -17.35 & 0.01 & -18.43 & 0.00 & -18.45 & 0.01 \\ 
\hline
     \end{tabular}
 \end{table*}

 \begin{table}
     \centering
\caption{Derived light curve properties of SN 2023bee and the comparison SNe in the $J$, $H$, and $K_{\rm s}$ bands. The table lists the time difference between the first and secondary maximum ($\Delta t(\lambda)$).}
\label{tab:derived_properties}
     \begin{tabular}{lcccccc}
     \hline
Object & $\Delta$ $t_{J}$ & err & $\Delta$ $t_{H}$ & err & $\Delta$ $t_{Ks}$  & err \\
       & (day) &   & (day) &  & (day) &  \\
     \hline \hline
23bee & 38.39 & 1.51 & 32.78 & 0.78 & 28.12 & 2.05  \\ 
01el & 34.51 & 0.54 & 24.87 & 1.56 & 26.48 & 1.08  \\ 
05el & 27.66 & 0.56 & 23.52 & 1.46 & 23.85 & 0.86 \\ 
06X & 31.70 & 0.51 & 25.79 & 1.69 & 22.69 & 1.78  \\ 
07le & 37.87 & 0.32 & 32.38 & 2.40 & 29.93 & 1.37  \\ 
08hs & 19.33 & 0.99 & 17.73 & 1.64 & 13.97 & 2.84 \\ 
09an & 29.00 & 0.50 & 25.06 & 0.86 & 22.55 & 1.81 \\ 
11B & 32.81 & 0.39 & 24.69 & 1.61 & 23.41 & 2.10  \\ 
11ao & 37.51 & 2.71 & 34.54 & 2.17 & $\cdot\cdot\cdot$ & $\cdot\cdot\cdot$  \\ 
12fr & 40.19 & 3.77 & 33.72 & 1.03 & $\cdot\cdot\cdot$ & $\cdot\cdot\cdot$  \\ 
17cbv & 36.19 & 0.41 & 30.02 & 1.34 & 28.02 & 1.22  \\      
\hline
     \end{tabular}
 \end{table}

\begin{figure}
\includegraphics[width=0.45\textwidth]{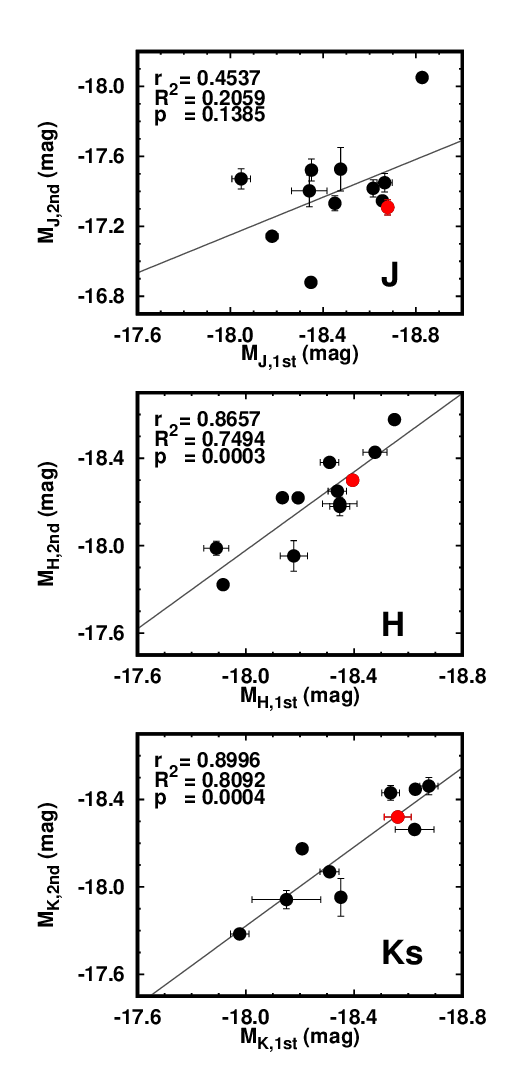}
\caption{Comparison of the absolute magnitudes at the first 
and secondary maxima for SN 2023bee and comparison SNe Ia. 
The $J$, $H$, and $K_{\rm s}$ bands are shown in the top, middle, and bottom panels, respectively. SN 2023bee is highlighted by a 
red filled circles. The solid lines indicate the regression line derived using the method of least squares. Here, $r$ represents the correlation coefficient, $R^2$ is the coefficient of determination, and $p$ denotes the probability value. {Alt text: Three scatter plots including regression lines.}}
\label{fig:abs_mag_detailed}
\end{figure}



\section{Discussion}

\subsection{Statistical analysis}

Here, we present a concise statistical analysis using the NIR light curves of SNe Ia, including SN 2023bee, and compare our methodology with the statistical research conducted by \citet{Dhawan2015}.
The key differences between our investigation and \citet{Dhawan2015} are as follows:
\begin{enumerate}
    \item We utilized $J$, $H$, and $K_{\rm s}$-band light curve data, whereas \citet{Dhawan2015} used $Y$, $J$, and $H$-band data.
    \item We determined the time difference between the first and the secondary maximum ($\Delta t$) using only the NIR light curves. In contrast, \citet{Dhawan2015} used the $B$-band maximum as the reference point for determining the secondary maximum date.
    \item Our comparison sample was strictly limited to SNe with data available from the pre-maximum phase until after the secondary maximum. The analysis in \citet{Dhawan2015} included samples lacking pre-maximum data.
\end{enumerate}
Consequently, the number of samples in our investigation is significantly smaller than that presented by \citet{Dhawan2015}. However, adhering to these criteria reduces the uncertainty in the derived maximum magnitudes and dates.

We performed light curve fitting to obtain essential properties, such as the first and second maximum absolute magnitudes ($M_{\lambda, \mathrm{1st}}$ and $M_{\lambda, \mathrm{2nd}}$), and the time difference between the two maxima ($\Delta t(\lambda)$).
We used a third-order polynomial function to fit the light curves and determine $M_{\lambda, \mathrm{1st}}$, $M_{\lambda, \mathrm{2nd}}$, and $\Delta t(\lambda)$.
The results of these derived properties are summarized in Tables~2 and ~3.
We analyzed the relationship between these parameters and calculated the correlation coefficient. To test its statistical significance, a t-test was performed to obtain the p-value.
The relationship between the first and second maximum magnitudes is presented in Figure~5.


In the $H$-band, the absolute magnitude of the first maximum shows a strong correlation with that of the secondary maximum, accompanied by a significant $p$-value ($p<0.05$). A similar trend is also found in the $K_{\rm s}$-band.
The correlation in the $K_{s}$-band is slightly stronger than that in the $H$-band.
In the $J$-band, there is no significant correlation between the absolute magnitudes of the first and secondary maxima. The secondary peak magnitudes remained nearly constant with an standard deviation of approximately 0.05 mag regardless of the first maximum magnitude, with the exceptions of SNe 2007le, 2008hs, and 2009an.

\begin{figure*}
\includegraphics[width=0.45\textwidth]{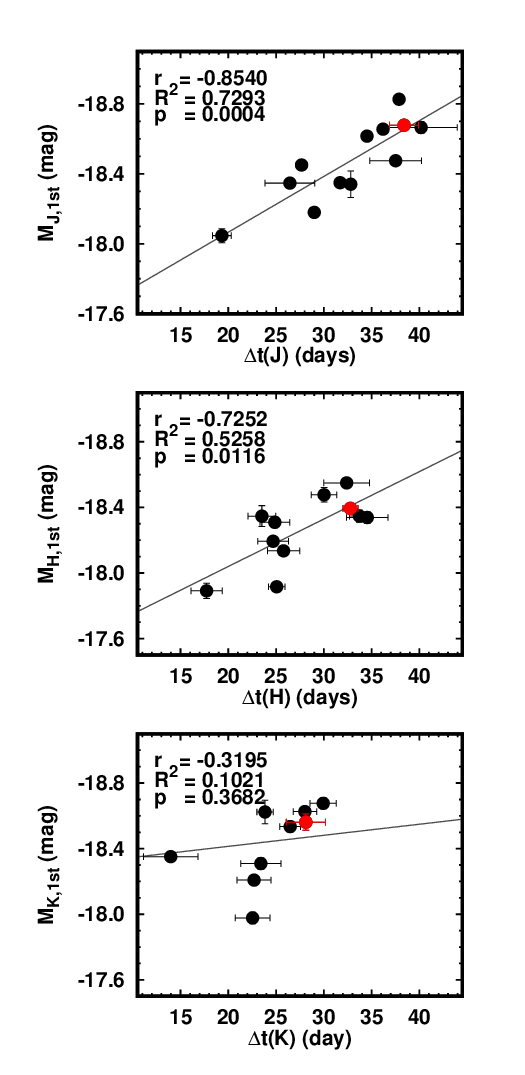} 
\includegraphics[width=0.45\textwidth]{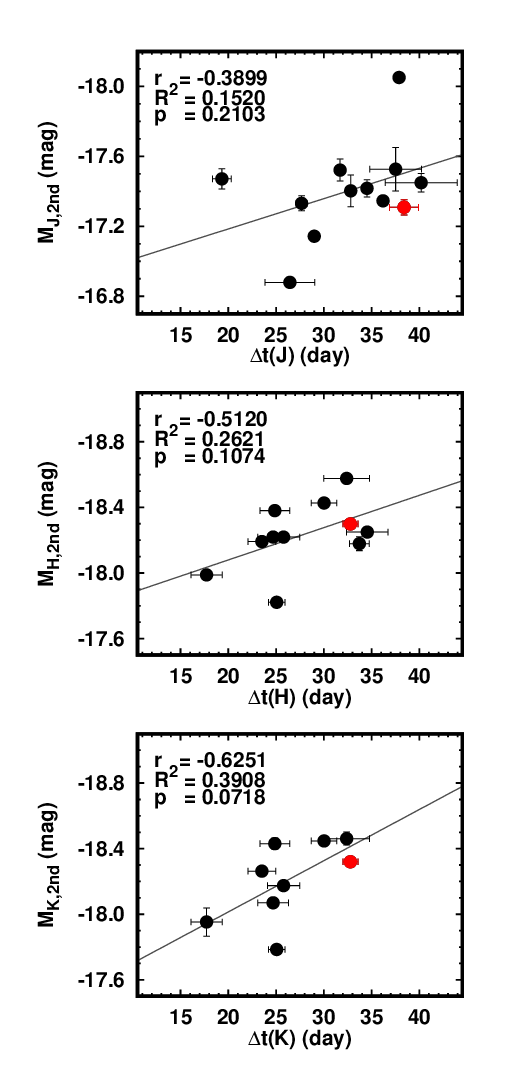}
\caption{The correlation plots between the absolute magnitudes and the time interval between maxima ($\Delta t(\lambda)$). The left three panels display the first maximum magnitudes ($M_{\lambda, 1\text{st}}$) versus $\Delta t(\lambda)$. The right panels show the secondary maximum magnitudes ($M_{\lambda, 2\text{nd}}$) versus $\Delta t(\lambda)$. 
{Alt text: Scatter plots arranged in left and right columns.}}
\label{fig:magnitude_vs_dt_short}
\end{figure*}

The inverse correlation between the first maximum absolute magnitude, $M_{J, 1st}$, and the time difference between the two peaks, $\Delta t(J)$, was found to be significant (see Figure~6).
A correlation between $M_{H, 1st}$ and $\Delta t(H)$ was also observed, but it was relatively weak.
In the $K_{\rm s}$ band, no significant correlation was found between $M_{K, 1st}$ and $\Delta t(K)$. 

No significant correlation was found between the secondary maximum magnitude, $M_{\text{2nd}}$, and $\Delta t$ for any of the $J$-, $H$-, and $K_{s}$-bands. In particular, the lack of correlation in the $J$-band directly reflects the observed constancy of $M_{J,\text{2nd}}$ across the sample.

\subsection{Explosion model interpretation}
\label{sec:explosion_model}

In our investigation, the $J$-band secondary peak magnitudes exhibit a relatively homogeneous distribution, with the exceptions of SNe 2007le, 2008hs and 2009an (upper panel of Figure~5). This similar trend was previously identified by \citet{Dhawan2015}.
This homogeneity suggests that the $J$-band luminosity is suppressed by absorption in the spectra of more luminous objects.
\citet{Hsiao2019} showed the NIR spectra of SN 2012fr in the $1.2\text{-}2.5\ \mu\mathrm{m}$ range at $t=16$, $31$, and $52$ d, where they observed multiple absorption features near $1.2\text{-}1.3\ \mu\mathrm{m}$. These features were identified by \citet{Taubenberger2011} as iron and cobalt lines in the extremely luminous SN Ia 2009dc.
This evidence supports the hypothesis that the absorption by iron-group elements becomes stronger in more luminous SNe Ia.
The constant $J$-band luminosity observed in normal and luminous SNe was also predicted by theoretical radiative transfer calculations based on the delayed detonation model \citep{Kasen2006}.

\citet{Kasen2006} presented theoretical NIR light curves based on the delayed detonation model. This model successfully reproduces the double-peaked light curve characteristic of SNe Ia because the excitation state of iron-group elements changes as the ejecta expands. Given the stratified structure of the ejecta, this change in the excitation state efficiently produces the double-peaked light curve. In this model, a larger explosion energy synthesizes a greater amount of $^{56}\mathrm{Ni}$ and stable Fe mass. This is consistent with the finding that SNe Ia showing a more luminous maximum also exhibit a later secondary maximum in the $J$ band.


\section{Conclusion}
\label{sec:conclusion}

Based on our analysis of the NIR photometric observations of the slow-declining SN Ia 2023bee, we draw the following conclusions.

     The $JHKs$-band light curves and the $J-H$ and $J-Ks$ color evolution of SN 2023bee are highly consistent with those of other slow-evolving SNe Ia, specifically SN 2012fr and SN 2017cbv. Previous spectroscopic studies \citep{Hosseinzadeh2023,QWang2024} classified SN 2023bee as a shallow-silicon SN Ia. The absolute magnitudes of the first and secondary maxima indicate that this supernova is located at the bright end of the luminosity distribution among SNe Ia, and its slow evolution is consistent with this spectroscopic classification.

    We obtained the NIR light curve properties for a sample of a dozen double-peaked SNe Ia and conducted a concise statistical investigation.
    \begin{itemize}
        \item The absolute magnitudes of the first and secondary maxima exhibit strong correlations in the $H$ and $K_s$ bands ($r = 0.8657$, $p = 0.0003$ and $r = 0.8996$, $p = 0.0004$, respectively).
        \item For the $J$ and $H$ bands, SNe Ia with a more luminous first maximum tend to exhibit a later secondary maximum, with correlation coefficients of $r=-0.854$ ($p=0.0004$) and $r=-0.7252$ ($p=0.0116$), respectively. In particular, the time between the first and secondary maxima in the $J$-band could serve as a potential indicator of the first maximum absolute magnitude. The correlation is weaker in the $H$ band than in the $J$ band.
        \item The $J$-band secondary maximum exhibits nearly constant luminosity, with a standard deviation of 0.05 mag, except for three outlying objects.
        This consistency aligns with the prediction of the delayed detonation model \citep{Kasen2006} and the strong absorption features of iron and cobalt seen in the NIR spectra of SNe like 2012fr \citep{Hsiao2019}.
    \end{itemize}

    Given the constraints derived from the $J$-band homogeneity, the investigation using NIR spectra is strongly encouraged.
    The samples used in this study are limited to objects that have 
    pre-maximum photometric points. To standardize the NIR light curves 
    of SNe Ia, it is essential to increase the number of 
    NIR light curve samples covering the period from before the first 
    maximum to after the secondary maximum.

    For future cosmological studies leveraging large survey observations 
    (e.g., those conducted by the \textit{Roman Space Telescope} \citep{Spergel2015} and \textit{Euclid} \citep{Laureijs2011}, a statistical study dedicated to the standardization of NIR light curves with a larger sample size is essential.

\begin{ack}
 We are grateful to graduate and undergraduate students for performing 
 the near-infrared observations. This work was supported by Grant-in-Aid 
 for Scientific Research (C) 22K03676. The Kagoshima University 1 m telescope is a member of the Optical and Infrared Synergetic Telescopes for Education and Research (OISTER) program funded by the MEXT of Japan. 
 The authors acknowledge the use of Gemini (Google AI) for English translation, editing, and stylistic improvements during the preparation of this manuscript. The authors retain full responsibility for the final content.
\end{ack}




\bibliography{addsample}{}

\begin{thebibliography}{}
\expandafter\ifx\csname natexlab\endcsname\relax\def\natexlab#1{#1}\fi

\bibitem[{{Andrews} {et~al.}(2023){Andrews}, {Sand}, {Valenti}, {Bostroem},
  {Wyatt}, {Lundquist}, {Hoang}, {Jencson}, {Paraskeva}, {Janzen}, {Shrestha},
  {Hosseinzadeh}, {Pearson}, {Dong}, \& {Meza}}]{Andrews2023}
{Andrews}, J.~E., {Sand}, D.~J., {Valenti}, S., {et~al.} 2023, Transient Name
  Server Discovery Report, 2023-256, 1

\bibitem[{{Arnett}(1982)}]{Arnett1982}
{Arnett}, W.~D. 1982, \apj, 253, 785

\bibitem[{{Avelino} {et~al.}(2019){Avelino}, {Friedman}, {Mandel}, {Jones},
  {Challis}, \& {Kirshner}}]{Avelino2019}
{Avelino}, A., {Friedman}, A.~S., {Mandel}, K.~S., {et~al.} 2019, \apj, 887,
  106

\bibitem[{{Axelrod}(1980)}]{Axelrod1980}
{Axelrod}, T.~S. 1980, PhD thesis, University of California, Santa Cruz

\bibitem[{{Barone-Nugent} {et~al.}(2012){Barone-Nugent}, {Lidman}, {Wyithe},
  {Mould}, {Howell}, {Hook}, {Sullivan}, {Nugent}, {Arcavi}, {Cenko}, {Cooke},
  {Gal-Yam}, {Hsiao}, {Kasliwal}, {Maguire}, {Ofek}, {Poznanski}, \&
  {Xu}}]{Barone-Nugent2012}
{Barone-Nugent}, R.~L., {Lidman}, C., {Wyithe}, J.~S.~B., {et~al.} 2012,
  \mnras, 425, 1007

\bibitem[{{Benetti} {et~al.}(2005){Benetti}, {Cappellaro}, {Mazzali},
  {Turatto}, {Altavilla}, {Bufano}, {Elias-Rosa}, {Kotak}, {Pignata}, {Salvo},
  \& {Stanishev}}]{Benetti2005}
{Benetti}, S., {Cappellaro}, E., {Mazzali}, P.~A., {et~al.} 2005, \apj, 623,
  1011

\bibitem[{{Branch} {et~al.}(1993){Branch}, {Fisher}, \& {Nugent}}]{Branch1993}
{Branch}, D., {Fisher}, A., \& {Nugent}, P. 1993, \aj, 106, 2383

\bibitem[{{Burns} {et~al.}(2020){Burns}, {Ashall}, {Contreras}, {Brown},
  {Stritzinger}, {Phillips}, {Flores}, {Suntzeff}, {Hsiao}, {Uddin}, {Simon},
  {Krisciunas}, {Campillay}, {Foley}, {Freedman}, {Galbany}, {Gonz{\'a}lez},
  {Hoeflich}, {Holmbo}, {Kilpatrick}, {Kirshner}, {Morrell},
  {Mu{\~n}oz-Elgueta}, {Piro}, {Rojas-Bravo}, {Sand}, {Vargas-Gonz{\'a}lez},
  {Ulloa}, \& {Vilchez}}]{Burns2020}
{Burns}, C.~R., {Ashall}, C., {Contreras}, C., {et~al.} 2020, \apj, 895, 118

\bibitem[{{Childress} {et~al.}(2013){Childress}, {Scalzo}, {Sim}, {Tucker},
  {Yuan}, {Schmidt}, {Cenko}, {Silverman}, {Contreras}, {Hsiao}, {Phillips},
  {Morrell}, {Jha}, {McCully}, {Filippenko}, {Anderson}, {Benetti}, {Bufano},
  {de Jaeger}, {Forster}, {Gal-Yam}, {Le Guillou}, {Maguire}, {Maund},
  {Mazzali}, {Pignata}, {Smartt}, {Spyromilio}, {Sullivan}, {Taddia},
  {Valenti}, {Bayliss}, {Bessell}, {Blanc}, {Carson}, {Clubb}, {de Burgh-Day},
  {Desjardins}, {Fang}, {Fox}, {Gates}, {Ho}, {Keller}, {Kelly}, {Lidman},
  {Loaring}, {Mould}, {Owers}, {Ozbilgen}, {Pei}, {Pickering}, {Pracy}, {Rich},
  {Schaefer}, {Scott}, {Stritzinger}, {Vogt}, \& {Zhou}}]{Childress2013}
{Childress}, M.~J., {Scalzo}, R.~A., {Sim}, S.~A., {et~al.} 2013, \apj, 770, 29

\bibitem[{{Contreras} {et~al.}(2018){Contreras}, {Phillips}, {Burns}, {Piro},
  {Shappee}, {Stritzinger}, {Baltay}, {Brown}, {Conseil}, {Klotz}, {Nugent},
  {Turpin}, {Parker}, {Rabinowitz}, {Hsiao}, {Morrell}, {Campillay},
  {Castell{\'o}n}, {Corco}, {Gonz{\'a}lez}, {Krisciunas}, {Ser{\'o}n},
  {Tucker}, {Walker}, {Baron}, {Cain}, {Childress}, {Folatelli}, {Freedman},
  {Hamuy}, {Hoeflich}, {Persson}, {Scalzo}, {Schmidt}, \&
  {Suntzeff}}]{Contreras2018}
{Contreras}, C., {Phillips}, M.~M., {Burns}, C.~R., {et~al.} 2018, \apj, 859,
  24

\bibitem[{{Deckers} {et~al.}(2023){Deckers}, {Graur}, {Maguire}, {Shingles},
  {Brennan}, {Anderson}, {Burke}, {Chen}, {Galbany}, {Grayling},
  {Guti{\'e}rrez}, {Harvey}, {Hiramatsu}, {Howell}, {Inserra}, {Killestein},
  {McCully}, {M{\"u}ller-Bravo}, {Nicholl}, {Newsome}, {Padilla Gonzalez},
  {Pellegrino}, {Terreran}, {Terwel}, {Toy}, \& {Young}}]{Deckers2023}
{Deckers}, M., {Graur}, O., {Maguire}, K., {et~al.} 2023, \mnras, 521, 4414

\bibitem[{{Dhawan} {et~al.}(2017){Dhawan}, {Leibundgut}, {Spyromilio}, \&
  {Blondin}}]{Dhawan2017}
{Dhawan}, S., {Leibundgut}, B., {Spyromilio}, J., \& {Blondin}, S. 2017, \aap,
  602, A118

\bibitem[{{Dhawan} {et~al.}(2015){Dhawan}, {Leibundgut}, {Spyromilio}, \&
  {Maguire}}]{Dhawan2015}
{Dhawan}, S., {Leibundgut}, B., {Spyromilio}, J., \& {Maguire}, K. 2015,
  \mnras, 448, 1345

\bibitem[{{Dhawan} {et~al.}(2023){Dhawan}, {Thorp}, {Mandel}, {Ward},
  {Narayan}, {Jha}, \& {Chant}}]{Dhawan2023}
{Dhawan}, S., {Thorp}, S., {Mandel}, K.~S., {et~al.} 2023, \mnras, 524, 235

\bibitem[{{Do} {et~al.}(2026){Do}, {Mandel}, {Shappee}, {Tully}, {Tonry},
  {Rubin}, {Jones}, {Dixon}, {de Jaeger}, {Scolnic}, {Peterson}, \&
  {Burns}}]{Do2026}
{Do}, A., {Mandel}, K.~S., {Shappee}, B.~J., {et~al.} 2026, arXiv e-prints,
  arXiv:2606.17142

\bibitem[{{Elias} {et~al.}(1981){Elias}, {Frogel}, {Hackwell}, \&
  {Persson}}]{Elias1981}
{Elias}, J.~H., {Frogel}, J.~A., {Hackwell}, J.~A., \& {Persson}, S.~E. 1981,
  \apjl, 251, L13

\bibitem[{{Elias} {et~al.}(1985){Elias}, {Matthews}, {Neugebauer}, \&
  {Persson}}]{Elias1985}
{Elias}, J.~H., {Matthews}, K., {Neugebauer}, G., \& {Persson}, S.~E. 1985,
  \apj, 296, 379

\bibitem[{{Elias-Rosa} {et~al.}(2008){Elias-Rosa}, {Benetti}, {Turatto},
  {Cappellaro}, {Valenti}, {Arkharov}, {Beckman}, {di Paola}, {Dolci},
  {Filippenko}, {Foley}, {Krisciunas}, {Larionov}, {Li}, {Meikle},
  {Pastorello}, {Valentini}, \& {Hillebrandt}}]{Elias-Rosa2008}
{Elias-Rosa}, N., {Benetti}, S., {Turatto}, M., {et~al.} 2008, \mnras, 384, 107

\bibitem[{{Filippenko} {et~al.}(1992){Filippenko}, {Richmond}, {Matheson},
  {Shields}, {Burbidge}, {Cohen}, {Dickinson}, {Malkan}, {Nelson}, {Pietz},
  {Schlegel}, {Schmeer}, {Spinrad}, {Steidel}, {Tran}, \&
  {Wren}}]{Filippenko1992b}
{Filippenko}, A.~V., {Richmond}, M.~W., {Matheson}, T., {et~al.} 1992, \apjl,
  384, L15

\bibitem[{{Friedman} {et~al.}(2015){Friedman}, {Wood-Vasey}, {Marion},
  {Challis}, {Mandel}, {Bloom}, {Modjaz}, {Narayan}, {Hicken}, {Foley},
  {Klein}, {Starr}, {Morgan}, {Rest}, {Blake}, {Miller}, {Falco}, {Wyatt},
  {Mink}, {Skrutskie}, \& {Kirshner}}]{Friedman2015}
{Friedman}, A.~S., {Wood-Vasey}, W.~M., {Marion}, G.~H., {et~al.} 2015, \apjs,
  220, 9

\bibitem[{{Garavini} {et~al.}(2004){Garavini}, {Folatelli}, {Goobar}, {Nobili},
  {Aldering}, {Amadon}, {Amanullah}, {Astier}, {Balland}, {Blanc}, {Burns},
  {Conley}, {Dahl{\'e}n}, {Deustua}, {Ellis}, {Fabbro}, {Fan}, {Frye}, {Gates},
  {Gibbons}, {Goldhaber}, {Goldman}, {Groom}, {Haissinski}, {Hardin}, {Hook},
  {Howell}, {Kasen}, {Kent}, {Kim}, {Knop}, {Lee}, {Lidman}, {Mendez},
  {Miller}, {Moniez}, {Mour{\~a}o}, {Newberg}, {Nugent}, {Pain}, {Perdereau},
  {Perlmutter}, {Prasad}, {Quimby}, {Raux}, {Regnault}, {Rich}, {Richards},
  {Ruiz-Lapuente}, {Sainton}, {Schaefer}, {Schahmaneche}, {Smith}, {Spadafora},
  {Stanishev}, {Walton}, {Wang}, {Wood-Vasey}, \& {Supernova Cosmology
  Project}}]{Garavini2004}
{Garavini}, G., {Folatelli}, G., {Goobar}, A., {et~al.} 2004, \aj, 128, 387

\bibitem[{{Graur} {et~al.}(2020){Graur}, {Maguire}, {Ryan}, {Nicholl},
  {Avelino}, {Riess}, {Shingles}, {Seitenzahl}, \& {Fisher}}]{Graur2020}
{Graur}, O., {Maguire}, K., {Ryan}, R., {et~al.} 2020, Nature Astronomy, 4, 188

\bibitem[{{Graur} {et~al.}(2023){Graur}, {Padilla Gonzalez}, {Burke},
  {Deckers}, {Jha}, {Galbany}, {Karamehmetoglu}, {Stritzinger}, {Maguire},
  {Howell}, {Fisher}, {Fullard}, {Handberg}, {Hiramatsu}, {Hosseinzadeh},
  {Kerzendorf}, {McCully}, {Newsome}, {Pellegrino}, {Rest}, {Riess},
  {Seitenzahl}, {Shara}, {Shen}, {Terreran}, \& {Zurek}}]{Graur2023}
{Graur}, O., {Padilla Gonzalez}, E., {Burke}, J., {et~al.} 2023, \mnras, 526,
  2977

\bibitem[{{Hachinger} {et~al.}(2006){Hachinger}, {Mazzali}, \&
  {Benetti}}]{Hachinger2006}
{Hachinger}, S., {Mazzali}, P.~A., \& {Benetti}, S. 2006, \mnras, 370, 299

\bibitem[{{Hamuy} {et~al.}(2006){Hamuy}, {Folatelli}, {Morrell}, {Phillips},
  {Suntzeff}, {Persson}, {Roth}, {Gonzalez}, {Krzeminski}, {Contreras},
  {Freedman}, {Murphy}, {Madore}, {Wyatt}, {Maza}, {Filippenko}, {Li}, \&
  {Pinto}}]{Hamuy2006}
{Hamuy}, M., {Folatelli}, G., {Morrell}, N.~I., {et~al.} 2006, \pasp, 118, 2

\bibitem[{{Hatano} {et~al.}(2000){Hatano}, {Branch}, {Lentz}, {Baron},
  {Filippenko}, \& {Garnavich}}]{Hatano2000}
{Hatano}, K., {Branch}, D., {Lentz}, E.~J., {et~al.} 2000, \apjl, 543, L49

\bibitem[{{Hosseinzadeh} {et~al.}(2023{\natexlab{a}}){Hosseinzadeh}, {Newsome},
  {Sand}, {Farah}, {Howell}, {McCully}, {Gonzalez}, {Pellegrino}, \&
  {Terreran}}]{Hosseinzadeh2023b}
{Hosseinzadeh}, G., {Newsome}, M., {Sand}, D.~J., {et~al.} 2023{\natexlab{a}},
  Transient Name Server Classification Report, 2023-277, 1

\bibitem[{{Hosseinzadeh} {et~al.}(2023{\natexlab{b}}){Hosseinzadeh}, {Sand},
  {Sarbadhicary}, {Ryder}, {Jha}, {Dong}, {Bostroem}, {Andrews}, {Hoang},
  {Janzen}, {Jencson}, {Lundquist}, {Meza Retamal}, {Pearson}, {Shrestha},
  {Valenti}, {Wyatt}, {Farah}, {Howell}, {McCully}, {Newsome}, {Padilla
  Gonzalez}, {Pellegrino}, {Terreran}, {Alzaabi}, {Green}, {Gurney}, {Milne},
  {Ridenhour}, {Smith}, {Robles}, {Kwok}, {Schwab}, {Gromadzki}, {Buckley},
  {Itagaki}, {Hiramatsu}, {Chomiuk}, {Lundqvist}, {Haislip}, {Kouprianov}, \&
  {Reichart}}]{Hosseinzadeh2023}
{Hosseinzadeh}, G., {Sand}, D.~J., {Sarbadhicary}, S.~K., {et~al.}
  2023{\natexlab{b}}, \apjl, 953, L15

\bibitem[{{Hoyle} \& {Fowler}(1960)}]{Hoyle1960}
{Hoyle}, F., \& {Fowler}, W.~A. 1960, \apj, 132, 565

\bibitem[{{Hsiao} {et~al.}(2019){Hsiao}, {Phillips}, {Marion}, {Kirshner},
  {Morrell}, {Sand}, {Burns}, {Contreras}, {Hoeflich}, {Stritzinger},
  {Valenti}, {Anderson}, {Ashall}, {Baltay}, {Baron}, {Banerjee}, {Davis},
  {Diamond}, {Folatelli}, {Freedman}, {F{\"o}rster}, {Galbany}, {Gall},
  {Gonz{\'a}lez-Gait{\'a}n}, {Goobar}, {Hamuy}, {Holmbo}, {Kasliwal},
  {Krisciunas}, {Kumar}, {Lidman}, {Lu}, {Nugent}, {Perlmutter}, {Persson},
  {Piro}, {Rabinowitz}, {Roth}, {Ryder}, {Schmidt}, {Shahbandeh}, {Suntzeff},
  {Taddia}, {Uddin}, \& {Wang}}]{Hsiao2019}
{Hsiao}, E.~Y., {Phillips}, M.~M., {Marion}, G.~H., {et~al.} 2019, \pasp, 131,
  014002

\bibitem[{{Iben} \& {Tutukov}(1984)}]{Iben1984}
{Iben}, I., J., \& {Tutukov}, A.~V. 1984, \apjs, 54, 335

\bibitem[{{Jack} {et~al.}(2015){Jack}, {Baron}, \& {Hauschildt}}]{Jack2015}
{Jack}, D., {Baron}, E., \& {Hauschildt}, P.~H. 2015, \mnras, 449, 3581

\bibitem[{{Johansson} {et~al.}(2021){Johansson}, {Cenko}, {Fox}, {Dhawan},
  {Goobar}, {Stanishev}, {Butler}, {Lee}, {Watson}, {Fremling}, {Kasliwal},
  {Nugent}, {Petrushevska}, {Sollerman}, {Yan}, {Burke}, {Hosseinzadeh},
  {Howell}, {McCully}, \& {Valenti}}]{Johansson2021}
{Johansson}, J., {Cenko}, S.~B., {Fox}, O.~D., {et~al.} 2021, \apj, 923, 237

\bibitem[{{Jones} {et~al.}(2022){Jones}, {Mandel}, {Kirshner}, {Thorp},
  {Challis}, {Avelino}, {Brout}, {Burns}, {Foley}, {Pan}, {Scolnic}, {Siebert},
  {Chornock}, {Freedman}, {Friedman}, {Frieman}, {Galbany}, {Hsiao}, {Kelsey},
  {Marion}, {Nichol}, {Nugent}, {Phillips}, {Rest}, {Riess}, {Sako}, {Smith},
  {Wiseman}, \& {Wood-Vasey}}]{Jones2022}
{Jones}, D.~O., {Mandel}, K.~S., {Kirshner}, R.~P., {et~al.} 2022, \apj, 933,
  172

\bibitem[{{Kasen}(2006)}]{Kasen2006}
{Kasen}, D. 2006, \apj, 649, 939

\bibitem[{{Kasen} {et~al.}(2009){Kasen}, {R{\"o}pke}, \& {Woosley}}]{Kasen2009}
{Kasen}, D., {R{\"o}pke}, F.~K., \& {Woosley}, S.~E. 2009, \nat, 460, 869

\bibitem[{{Khokhlov}(1991)}]{Khokhlov1991}
{Khokhlov}, A.~M. 1991, \aap, 245, 114

\bibitem[{{Krisciunas} {et~al.}(2004){Krisciunas}, {Phillips}, \&
  {Suntzeff}}]{Krisciunas2004}
{Krisciunas}, K., {Phillips}, M.~M., \& {Suntzeff}, N.~B. 2004, \apjl, 602, L81

\bibitem[{{Krisciunas} {et~al.}(2001){Krisciunas}, {Phillips}, {Stubbs},
  {Rest}, {Miknaitis}, {Riess}, {Suntzeff}, {Roth}, {Persson}, \&
  {Freedman}}]{Krisciunas2001}
{Krisciunas}, K., {Phillips}, M.~M., {Stubbs}, C., {et~al.} 2001, \aj, 122,
  1616

\bibitem[{{Krisciunas} {et~al.}(2003){Krisciunas}, {Suntzeff}, {Candia},
  {Arenas}, {Espinoza}, {Gonzalez}, {Gonzalez}, {H{\"o}flich}, {Landolt},
  {Phillips}, \& {Pizarro}}]{Krisciunas2003}
{Krisciunas}, K., {Suntzeff}, N.~B., {Candia}, P., {et~al.} 2003, \aj, 125, 166

\bibitem[{{Laureijs} {et~al.}(2011){Laureijs}, {Amiaux}, {Arduini},
  {Augu{\`e}res}, {Brinchmann}, {Cole}, {Cropper}, {Dabin}, {Duvet}, {Ealet},
  {Garilli}, {Gondoin}, {Guzzo}, {Hoar}, {Hoekstra}, {Holmes}, {Kitching},
  {Maciaszek}, {Mellier}, {Pasian}, {Percival}, {Rhodes}, {Saavedra Criado},
  {Sauvage}, {Scaramella}, {Valenziano}, {Warren}, {Bender}, {Castander},
  {Cimatti}, {Le F{\`e}vre}, {Kurki-Suonio}, {Levi}, {Lilje}, {Meylan},
  {Nichol}, {Pedersen}, {Popa}, {Rebolo Lopez}, {Rix}, {Rottgering},
  {Zeilinger}, {Grupp}, {Hudelot}, {Massey}, {Meneghetti}, {Miller}, {Paltani},
  {Paulin-Henriksson}, {Pires}, {Saxton}, {Schrabback}, {Seidel}, {Walsh},
  {Aghanim}, {Amendola}, {Bartlett}, {Baccigalupi}, {Beaulieu}, {Benabed},
  {Cuby}, {Elbaz}, {Fosalba}, {Gavazzi}, {Helmi}, {Hook}, {Irwin}, {Kneib},
  {Kunz}, {Mannucci}, {Moscardini}, {Tao}, {Teyssier}, {Weller}, {Zamorani},
  {Zapatero Osorio}, {Boulade}, {Foumond}, {Di Giorgio}, {Guttridge}, {James},
  {Kemp}, {Martignac}, {Spencer}, {Walton}, {Bl{\"u}mchen}, {Bonoli},
  {Bortoletto}, {Cerna}, {Corcione}, {Fabron}, {Jahnke}, {Ligori}, {Madrid},
  {Martin}, {Morgante}, {Pamplona}, {Prieto}, {Riva}, {Toledo}, {Trifoglio},
  {Zerbi}, {Abdalla}, {Douspis}, {Grenet}, {Borgani}, {Bouwens}, {Courbin},
  {Delouis}, {Dubath}, {Fontana}, {Frailis}, {Grazian}, {Koppenh{\"o}fer},
  {Mansutti}, {Melchior}, {Mignoli}, {Mohr}, {Neissner}, {Noddle}, {Poncet},
  {Scodeggio}, {Serrano}, {Shane}, {Starck}, {Surace}, {Taylor},
  {Verdoes-Kleijn}, {Vuerli}, {Williams}, {Zacchei}, {Altieri}, {Escudero
  Sanz}, {Kohley}, {Oosterbroek}, {Astier}, {Bacon}, {Bardelli}, {Baugh},
  {Bellagamba}, {Benoist}, {Bianchi}, {Biviano}, {Branchini}, {Carbone},
  {Cardone}, {Clements}, {Colombi}, {Conselice}, {Cresci}, {Deacon}, {Dunlop},
  {Fedeli}, {Fontanot}, {Franzetti}, {Giocoli}, {Garcia-Bellido}, {Gow},
  {Heavens}, {Hewett}, {Heymans}, {Holland}, {Huang}, {Ilbert}, {Joachimi},
  {Jennins}, {Kerins}, {Kiessling}, {Kirk}, {Kotak}, {Krause}, {Lahav}, {van
  Leeuwen}, {Lesgourgues}, {Lombardi}, {Magliocchetti}, {Maguire}, {Majerotto},
  {Maoli}, {Marulli}, {Maurogordato}, {McCracken}, {McLure}, {Melchiorri},
  {Merson}, {Moresco}, {Nonino}, {Norberg}, {Peacock}, {Pello}, {Penny},
  {Pettorino}, {Di Porto}, {Pozzetti}, {Quercellini}, {Radovich}, {Rassat},
  {Roche}, {Ronayette}, \& {Rossetti}}]{Laureijs2011}
{Laureijs}, R., {Amiaux}, J., {Arduini}, S., {et~al.} 2011, arXiv e-prints,
  arXiv:1110.3193

\bibitem[{{Lira} {et~al.}(1998){Lira}, {Suntzeff}, {Phillips}, {Hamuy}, {Maza},
  {Schommer}, {Smith}, {Wells}, {Avil{\'e}s}, {Baldwin}, {Elias},
  {Gonz{\'a}lez}, {Layden}, {Navarrete}, {Ugarte}, {Walker}, {Williger},
  {Baganoff}, {Crotts}, {Rich}, {Tyson}, {Dey}, {Guhathakurta}, {Hibbard},
  {Kim}, {Rehner}, {Siciliano}, {Roth}, {Seitzer}, \& {Williams}}]{Lira1998}
{Lira}, P., {Suntzeff}, N.~B., {Phillips}, M.~M., {et~al.} 1998, \aj, 115, 234

\bibitem[{{Maeda} {et~al.}(2015){Maeda}, {Nozawa}, {Nagao}, \&
  {Motohara}}]{Maeda2015}
{Maeda}, K., {Nozawa}, T., {Nagao}, T., \& {Motohara}, K. 2015, \mnras, 452,
  3281

\bibitem[{{Mandel} {et~al.}(2022){Mandel}, {Thorp}, {Narayan}, {Friedman}, \&
  {Avelino}}]{Mandel2022}
{Mandel}, K.~S., {Thorp}, S., {Narayan}, G., {Friedman}, A.~S., \& {Avelino},
  A. 2022, \mnras, 510, 3939

\bibitem[{{Meikle}(2000)}]{Meikle2000}
{Meikle}, W.~P.~S. 2000, \mnras, 314, 782

\bibitem[{{Nagao} {et~al.}(2017){Nagao}, {Maeda}, \& {Yamanaka}}]{Nagao2017}
{Nagao}, T., {Maeda}, K., \& {Yamanaka}, M. 2017, \apj, 835, 143

\bibitem[{{Nagayama} \& {Nakaya}(2024)}]{Nagayama2024}
{Nagayama}, T., \& {Nakaya}, H. 2024, in Society of Photo-Optical
  Instrumentation Engineers (SPIE) Conference Series, Vol. 13096, Ground-based
  and Airborne Instrumentation for Astronomy X, ed. J.~J. {Bryant},
  K.~{Motohara}, \& J.~R.~D. {Vernet}, 130963I

\bibitem[{{Nomoto}(1982)}]{Nomoto1982}
{Nomoto}, K. 1982, \apj, 253, 798

\bibitem[{{Nomoto} {et~al.}(1984){Nomoto}, {Thielemann}, \&
  {Yokoi}}]{Nomoto1984}
{Nomoto}, K., {Thielemann}, F., \& {Yokoi}, K. 1984, \apj, 286, 644

\bibitem[{{Nugent} {et~al.}(1995){Nugent}, {Phillips}, {Baron}, {Branch}, \&
  {Hauschildt}}]{Nugent1995}
{Nugent}, P., {Phillips}, M., {Baron}, E., {Branch}, D., \& {Hauschildt}, P.
  1995, \apjl, 455, L147

\bibitem[{{Perlmutter} {et~al.}(1999){Perlmutter}, {Aldering}, {Goldhaber},
  {Knop}, {Nugent}, {Castro}, {Deustua}, {Fabbro}, {Goobar}, {Groom}, {Hook},
  {Kim}, {Kim}, {Lee}, {Nunes}, {Pain}, {Pennypacker}, {Quimby}, {Lidman},
  {Ellis}, {Irwin}, {McMahon}, {Ruiz-Lapuente}, {Walton}, {Schaefer}, {Boyle},
  {Filippenko}, {Matheson}, {Fruchter}, {Panagia}, {Newberg}, {Couch}, \& {The
  Supernova Cosmology Project}}]{Perlmutter1999}
{Perlmutter}, S., {Aldering}, G., {Goldhaber}, G., {et~al.} 1999, \apj, 517,
  565

\bibitem[{{Peterson} {et~al.}(2023){Peterson}, {Jones}, {Scolnic},
  {S{\'a}nchez}, {Do}, {Riess}, {Ward}, {Dwomoh}, {de Jaeger}, {Jha}, {Mandel},
  {Pierel}, {Popovic}, {Rose}, {Rubin}, {Shappee}, {Thorp}, {Tonry}, {Tully},
  \& {Vincenzi}}]{Peterson2023}
{Peterson}, E.~R., {Jones}, D.~O., {Scolnic}, D., {et~al.} 2023, \mnras, 522,
  2478

\bibitem[{{Phillips}(1993)}]{Phillips1993}
{Phillips}, M.~M. 1993, \apjl, 413, L105

\bibitem[{{Phillips}(2012)}]{Phillips2012}
---. 2012, Publ. Astron. Soc. Australia, 29, 434

\bibitem[{{Phillips} {et~al.}(1999){Phillips}, {Lira}, {Suntzeff}, {Schommer},
  {Hamuy}, \& {Maza}}]{Phillips1999}
{Phillips}, M.~M., {Lira}, P., {Suntzeff}, N.~B., {et~al.} 1999, \aj, 118, 1766

\bibitem[{{Phillips} {et~al.}(1992){Phillips}, {Wells}, {Suntzeff}, {Hamuy},
  {Leibundgut}, {Kirshner}, \& {Foltz}}]{Phillips1992}
{Phillips}, M.~M., {Wells}, L.~A., {Suntzeff}, N.~B., {et~al.} 1992, \aj, 103,
  1632

\bibitem[{{Pisano} {et~al.}(2002){Pisano}, {Wilcots}, \& {Liu}}]{Pisano2002}
{Pisano}, D.~J., {Wilcots}, E.~M., \& {Liu}, C.~T. 2002, \apjs, 142, 161

\bibitem[{{Riess} {et~al.}(1998){Riess}, {Filippenko}, {Challis},
  {Clocchiatti}, {Diercks}, {Garnavich}, {Gilliland}, {Hogan}, {Jha},
  {Kirshner}, {Leibundgut}, {Phillips}, {Reiss}, {Schmidt}, {Schommer},
  {Smith}, {Spyromilio}, {Stubbs}, {Suntzeff}, \& {Tonry}}]{Riess1998}
{Riess}, A.~G., {Filippenko}, A.~V., {Challis}, P., {et~al.} 1998, \aj, 116,
  1009

\bibitem[{{Sadakane} {et~al.}(1996){Sadakane}, {Yokoo}, {Arimoto}, {Matsumoto},
  {Honda}, {Tanabe}, {Wakamatsu}, {Nishida}, {Yoshida}, \&
  {Takada-Hidai}}]{Sadakane1996}
{Sadakane}, K., {Yokoo}, T., {Arimoto}, J.-I., {et~al.} 1996, \pasj, 48, 51

\bibitem[{{Scalzo} {et~al.}(2014){Scalzo}, {Aldering}, {Antilogus}, {Aragon},
  {Bailey}, {Baltay}, {Bongard}, {Buton}, {Cellier-Holzem}, {Childress},
  {Chotard}, {Copin}, {Fakhouri}, {Gangler}, {Guy}, {Kim}, {Kowalski},
  {Kromer}, {Nordin}, {Nugent}, {Paech}, {Pain}, {Pecontal}, {Pereira},
  {Perlmutter}, {Rabinowitz}, {Rigault}, {Runge}, {Saunders}, {Sim}, {Smadja},
  {Tao}, {Taubenberger}, {Thomas}, {Weaver}, \& {Nearby Supernova
  Factory}}]{Scalzo2014}
{Scalzo}, R., {Aldering}, G., {Antilogus}, P., {et~al.} 2014, \mnras, 440, 1498

\bibitem[{{Schlafly} \& {Finkbeiner}(2011)}]{Schlafly2011}
{Schlafly}, E.~F., \& {Finkbeiner}, D.~P. 2011, \apj, 737, 103

\bibitem[{{Simon} {et~al.}(2009){Simon}, {Gal-Yam}, {Gnat}, {Quimby},
  {Ganeshalingam}, {Silverman}, {Blondin}, {Li}, {Filippenko}, {Wheeler},
  {Kirshner}, {Patat}, {Nugent}, {Foley}, {Vogt}, {Butler}, {Peek},
  {Rosolowsky}, {Herczeg}, {Sauer}, \& {Mazzali}}]{Simon2009}
{Simon}, J.~D., {Gal-Yam}, A., {Gnat}, O., {et~al.} 2009, \apj, 702, 1157

\bibitem[{{Skrutskie} {et~al.}(2006){Skrutskie}, {Cutri}, {Stiening},
  {Weinberg}, {Schneider}, {Carpenter}, {Beichman}, {Capps}, {Chester},
  {Elias}, {Huchra}, {Liebert}, {Lonsdale}, {Monet}, {Price}, {Seitzer},
  {Jarrett}, {Kirkpatrick}, {Gizis}, {Howard}, {Evans}, {Fowler}, {Fullmer},
  {Hurt}, {Light}, {Kopan}, {Marsh}, {McCallon}, {Tam}, {Van Dyk}, \&
  {Wheelock}}]{Skrutskie2006}
{Skrutskie}, M.~F., {Cutri}, R.~M., {Stiening}, R., {et~al.} 2006, \aj, 131,
  1163

\bibitem[{{Spergel} {et~al.}(2015){Spergel}, {Gehrels}, {Baltay}, {Bennett},
  {Breckinridge}, {Donahue}, {Dressler}, {Gaudi}, {Greene}, {Guyon}, {Hirata},
  {Kalirai}, {Kasdin}, {Macintosh}, {Moos}, {Perlmutter}, {Postman},
  {Rauscher}, {Rhodes}, {Wang}, {Weinberg}, {Benford}, {Hudson}, {Jeong},
  {Mellier}, {Traub}, {Yamada}, {Capak}, {Colbert}, {Masters}, {Penny},
  {Savransky}, {Stern}, {Zimmerman}, {Barry}, {Bartusek}, {Carpenter}, {Cheng},
  {Content}, {Dekens}, {Demers}, {Grady}, {Jackson}, {Kuan}, {Kruk}, {Melton},
  {Nemati}, {Parvin}, {Poberezhskiy}, {Peddie}, {Ruffa}, {Wallace}, {Whipple},
  {Wollack}, \& {Zhao}}]{Spergel2015}
{Spergel}, D., {Gehrels}, N., {Baltay}, C., {et~al.} 2015, arXiv e-prints,
  arXiv:1503.03757

\bibitem[{{Stetson}(1987)}]{Stetson1987}
{Stetson}, P.~B. 1987, \pasp, 99, 191

\bibitem[{{Taubenberger} {et~al.}(2011){Taubenberger}, {Benetti}, {Childress},
  {Pakmor}, {Hachinger}, {Mazzali}, {Stanishev}, {Elias-Rosa}, {Agnoletto},
  {Bufano}, {Ergon}, {Harutyunyan}, {Inserra}, {Kankare}, {Kromer},
  {Navasardyan}, {Nicolas}, {Pastorello}, {Prosperi}, {Salgado}, {Sollerman},
  {Stritzinger}, {Turatto}, {Valenti}, \& {Hillebrandt}}]{Taubenberger2011}
{Taubenberger}, S., {Benetti}, S., {Childress}, M., {et~al.} 2011, \mnras, 61

\bibitem[{{Tutukov} \& {Yungelson}(1979)}]{Tutukov1979}
{Tutukov}, A.~V., \& {Yungelson}, L.~R. 1979, \actaa, 29, 665

\bibitem[{{Wang} {et~al.}(2024){Wang}, {Rest}, {Dimitriadis}, {Ridden-Harper},
  {Siebert}, {Magee}, {Angus}, {Auchettl}, {Davis}, {Foley}, {Fox}, {Gomez},
  {Jencson}, {Jones}, {Kilpatrick}, {Pierel}, {Piro}, {Polin}, {Politsch},
  {Rojas-Bravo}, {Shahbandeh}, {Villar}, {Zenati}, {Ashall}, {Chambers},
  {Coulter}, {de Boer}, {DiLullo}, {Gall}, {Gao}, {Hsiao}, {Huber}, {Izzo},
  {Khetan}, {LeBaron}, {Magnier}, {Mandel}, {McGill}, {Miao}, {Pan}, {Stevens},
  {Swift}, {Taggart}, \& {Yang}}]{QWang2024}
{Wang}, Q., {Rest}, A., {Dimitriadis}, G., {et~al.} 2024, \apj, 962, 17

\bibitem[{{Wang} {et~al.}(2008){Wang}, {Li}, {Filippenko}, {Krisciunas},
  {Suntzeff}, {Li}, {Zhang}, {Deng}, {Foley}, {Ganeshalingam}, {Li}, {Lou},
  {Qiu}, {Shang}, {Silverman}, {Zhang}, \& {Zhang}}]{XWang2008}
{Wang}, X., {Li}, W., {Filippenko}, A.~V., {et~al.} 2008, \apj, 675, 626

\bibitem[{{Wee} {et~al.}(2018){Wee}, {Chakraborty}, {Wang}, \&
  {Penprase}}]{Wee2018}
{Wee}, J., {Chakraborty}, N., {Wang}, J., \& {Penprase}, B.~E. 2018, \apj, 863,
  90

\bibitem[{{Wells} {et~al.}(1994){Wells}, {Phillips}, {Suntzeff}, {Heathcote},
  {Hamuy}, {Navarrete}, {Fernandez}, {Weller}, {Schommer}, {Kirshner},
  {Leibundgut}, {Willner}, {Peletier}, {Schlegel}, {Wheeler}, {Harkness},
  {Bell}, {Matthews}, {Filippenko}, {Shields}, {Richmond}, {Jewitt}, {Luu},
  {Tran}, {Appleton}, {Robson}, {Tyson}, {Guhathakurta}, {Eder}, {Bond},
  {Potter}, {Veilleux}, {Porter}, {Humphreys}, {Janes}, {Williams}, {Costa},
  {Ruiz}, {Lee}, {Lutz}, {Rich}, {Winkler}, \& {Tyson}}]{Wells1994}
{Wells}, L.~A., {Phillips}, M.~M., {Suntzeff}, B., {et~al.} 1994, \aj, 108,
  2233

\bibitem[{{Wood-Vasey} {et~al.}(2007){Wood-Vasey}, {Miknaitis}, {Stubbs},
  {Jha}, {Riess}, {Garnavich}, {Kirshner}, {Aguilera}, {Becker}, {Blackman},
  {Blondin}, {Challis}, {Clocchiatti}, {Conley}, {Covarrubias}, {Davis},
  {Filippenko}, {Foley}, {Garg}, {Hicken}, {Krisciunas}, {Leibundgut}, {Li},
  {Matheson}, {Miceli}, {Narayan}, {Pignata}, {Prieto}, {Rest}, {Salvo},
  {Schmidt}, {Smith}, {Sollerman}, {Spyromilio}, {Tonry}, {Suntzeff}, \&
  {Zenteno}}]{Wood-Vasey2007}
{Wood-Vasey}, W.~M., {Miknaitis}, G., {Stubbs}, C.~W., {et~al.} 2007, \apj,
  666, 694

\bibitem[{{Yamanaka} {et~al.}(2016){Yamanaka}, {Maeda}, {Tanaka}, {Tominaga},
  {Kawabata}, {Takaki}, {Kawabata}, {Nakaoka}, {Ueno}, {Akitaya}, {Nagayama},
  {Takahashi}, {Honda}, {Omodaka}, {Miyanoshita}, {Nagao}, {Watanabe},
  {Isogai}, {Arai}, {Itoh}, {Ui}, {Uemura}, {Yoshida}, {Hanayama}, {Kuroda},
  {Ukita}, {Yanagisawa}, {Izumiura}, {Saito}, {Masumoto}, {Ono}, {Noguchi},
  {Matsumoto}, {Nogami}, {Morokuma}, {Oasa}, \& {Sekiguchi}}]{Yamanaka2016}
{Yamanaka}, M., {Maeda}, K., {Tanaka}, M., {et~al.} 2016, \pasj, 68, 68

\bibitem[{{Zhai} {et~al.}(2023){Zhai}, {Zhang}, {Li}, \& {Wang}}]{QZhai2023}
{Zhai}, Q., {Zhang}, J., {Li}, L., \& {Wang}, X. 2023, Transient Name Server
  Classification Report, 2023-260, 1

\end{thebibliography}
\bibliographystyle{apj}



\end{document}